\documentclass{aa}

\usepackage{times}
\usepackage{graphics}

\begin{document}

   \thesaurus{11(05.01.1, 04.03.1)}

   \title{The combination of ground-based astrometric compilation catalogues
          with the HIPPARCOS Catalogue}

   \subtitle{I. Single-star mode}

   \author{R. Wielen
   \and H. Lenhardt
   \and H. Schwan
   \and C. Dettbarn}

   \offprints{R. Wielen (wielen@ari.uni-heidelberg.de)}

   \institute{Astronomisches Rechen-Institut, Moenchhofstrasse 12-14,
   D-69120 Heidelberg, Germany}

   \date{Received 19 January 1999 / Accepted 20 April 1999}

\authorrunning{R. Wielen et al.}

\titlerunning
{Combination of ground-based astrometric compilation catalogues
with the HIPPARCOS Catalogue}

\maketitle

\begin{abstract}
The combination of ground-based astrometric compilation catalogues, such as the
FK5 or the GC, with the results of the ESA Astrometric Satellite HIPPARCOS
produces for many thousands of stars proper motions which are significantly
more accurate than the proper motions derived from the HIPPARCOS observations
alone. In the combination of the basic FK5 with the HIPPARCOS Catalogue (i.e.,
in the FK6), the gain in accuracy is about a factor of two for the proper
motions of single stars. The use of the GC still improves the accuracy of the
proper motions by a factor of about 1.2\,. We derive and describe
in detail how to combine a ground-based compilation
catalogue with HIPPARCOS. Our analytic approach is helpful for understanding
the principles of the combination method. In real applications we use a
numerical approach which avoids some (minor) approximations made in the
analytic approach. We give a numerical example of our combination method and
present an overall error budget for the combination of the ground-based
data for the basic FK5 stars and for the GC stars with the HIPPARCOS
observations.

In the present paper we describe the `single-star mode' of our combination
method. This mode is appropriate for truly single stars or for stars which can
be treated like single stars. The specific handling of binaries will be
discussed in subsequent papers.

\end{abstract}

%Section 1
\section{Introduction}

The HIPPARCOS astrometric satellite has provided
very accurate positions, proper motions, and parallaxes
for more than 118\,000 stars (ESA 1997).
Does this
mean that the long series of ground-based astrometric observations of all these
stars are now merely of historical value\,? This is certainly not the case: It
can be shown that the combination of the HIPPARCOS data with ground-based
results is providing for many thousands of stars individual proper motions
which are significantly more accurate than the HIPPARCOS proper motions
themselves (Wielen 1988, Wielen et al. 1998, 1999).

There are two main reasons why ground-based observations can improve  the
HIPPARCOS results: (1) For many of the
brighter stars, the ground-based observations
cover a period of time of more than two centuries. The measuring accuracy of
these ground-based positions is such that they allow, especially in combination
with the HIPPARCOS positions, to derive proper motions which have significantly
smaller measuring errors than the HIPPARCOS proper motions. It is well-known
that ground-based observations suffer from considerable systematic errors.
However, it is just the HIPPARCOS Catalogue itself which allows us to remove
these systematic errors from the ground-based data to such a high degree that
we can use with much confidence the corrected ground-based results for the
direct combination with HIPPARCOS. (2) The HIPPARCOS proper motions are derived
from observations which have been carried out within about three years only.
Since many of the HIPPARCOS stars are undetected astrometric binaries, these
nearly `instantaneously' measured HIPPARCOS proper motions can differ
significantly from `time-averaged', `mean' proper motions. We have called the
difference between the instantaneous HIPPARCOS proper motion and the mean
motion of a star the `cosmic error' of the HIPPARCOS proper motion (Wielen
1997). In many cases these cosmic errors are significantly larger than the
measuring errors of the instantaneous HIPPARCOS proper motions (Wielen 1995a,
b, Wielen et al. 1997, 1998, 1999).
The ground-based observations, with their
long observational history, are often already providing `mean' proper
motions. The ground-based data allow us therefore to identify and to correct
partially the cosmic errors in the HIPPARCOS proper motions.

Having argued that a combination of ground-based observations with the
HIPPARCOS data is useful for many stars, we now provide a method for
carrying out such a combination. The best way would be to combine the
individual observational catalogues of ground-based positions directly with
HIPPARCOS. However, there exists more than 2000 of such observational
catalogues. While we have collected most of these catalogues in our astrometric
data base ARIGFH (described in Wielen (1998)) and hope to use them
individually
in the future, we have to rely for the moment on a few compilation catalogues.
Such a compilation catalogue is usually based on a large number of
observational catalogues and provides as a `summary' of these individual
catalogues a derived proper motion and a mean position at a central epoch.

The most accurate of the ground-based compilation catalogues is the first part
of the FK5 (Fricke et al. 1988). It contains the 1535 basic fundamental stars.
The second part of the FK5 (Fricke et al. 1991) lists 3117 additional
stars in the bright and faint extension. A compilation of the remaining
FK4Sup stars (Schwan et al. 1993) provides data for further 995 stars with
a good observational history. The combination of the basic FK5 stars with
HIPPARCOS data produces the FK6. The proper motions of the FK6 are of
unprecedented accuracy (Wielen et al. 1998,1999, and Sect. 6).
The accuracy of
these FK6 proper motions rests to a high degree on the older observations. This
is not only to be expected on theoretical grounds, but can be
proved empirically  also
(Wielen et al. 1998) by combining HIPPARCOS with an older fundamental
catalogue such as the
FK3 (Kopff 1937, 1938). Because the old observations carry most of the weight
in a combination with HIPPARCOS, the GC (Boss et al. 1937) is also relevant for
our purpose. The GC is a careful compilation of ground-based observations from
the 18th century until about 1930. The main advantage of the GC is its large
number of stars. About 29\,700 of the 33\,342 GC stars have been observed
by HIPPARCOS and can therefore be used in our method. The typical accuracy
obtained for a GC star is, of course, lower than for an FK6 star (Wielen et al.
1998, 1999).

Instead of using a {\em compilation} catalogue, we may rely on one old
{\em observational} catalogue only, such as the Astrographic Catalogue (AC).
However, the formal accuracy of the proper motions obtained by a combination of
the AC with the HIPPARCOS data is in most cases only marginally better than
that of the HIPPARCOS proper motions alone. On the other hand, a comparison of
a proper motion derived from an AC position and a HIPPARCOS position with the
quasi-instantaneous HIPPARCOS proper motion $\mu_H$ can reveal the occurence of
a large value of the cosmic error in $\mu_H$ for the star under consideration.

Before we can combine the ground-based data with HIPPARCOS, the systematic
errors of the ground-based positions and proper motions have to be determined
and removed. For the determination of these systematic errors we use methods
developed for the construction of the FK5 (Bien et al. 1978). These methods do
not only provide the systematic errors themselves but also the local
uncertainty of the correction, i.e. the mean `measuring' error of the
systematic correction for a given star. As the reference system, we use the
HIPPARCOS system. The combined catalogue is therefore always on the HIPPARCOS
system, i.e. on the ICRS. For our purpose, we can consider the HIPPARCOS data
as free from systematic errors. A possible slight rotation of the HIPPARCOS
system with respect to an inertial, extragalactic system does not affect our
method (but of course our results).

In the following sections, we assume that the angular coordinates of a star,
$\alpha_\ast = \alpha\,\cos\,\delta$ and $\delta$,  change linearly in
time $t$. In practice, the non-linear motion of a star in $\alpha_\ast$ and
$\delta$ because of spherical effects and the foreshortening effect has to be
taken into account. We do this already in the determination of the systematic
errors between the ground-based data and HIPPARCOS. Later we do not use the
full values of ground-based positions and proper motions, but only their
differential values with respect to the HIPPARCOS solution at the proper epoch.
If we use these differential values only, the linear approximation is then
fully sufficient for our purpose.

We shall now present the mathematical details of our combination method. We
offer two approaches: an approximate analytic one and a rigorous
numerical one.
The analytic approach has the advantage to show more clearly the basic
principles. It neglects, however, some of the correlations which occur among
the HIPPARCOS results. In practical applications, we use the numerical approach
which takes all the correlation among the HIPPARCOS results into account. In
all cases, we assume that the HIPPARCOS  data are completely uncorrelated with
the ground-based data. This is certainly true for all of our stars, although
the HIPPARCOS Input Catalogue, used as a first approximation in the HIPPARCOS
data reduction procedure, had to rely on the ground-based observations.

%Section 2
\section{Analytic approach}

We assume that two astrometric catalogues are available. They are
identified by the indices 1 and 2. Examples are the FK5 and HIPPARCOS. For the
combined catalogue, e.g. the FK6, we use the index C. Each of the two basic
catalogues (i\,=\,1, 2) provides for all the stars under consideration a
position $x_i (T_i)$ and a proper motion $\mu_i$ at a central epoch $T_i$ for
two uncorrelated coordinate components, e.g. for $\alpha_\ast$ and $\delta$. In
addition, both catalogues provide mean errors of $x_i (T_i)$ and $\mu_i$ which
we denote by $\varepsilon_{x, i}$ and $\varepsilon_{\mu, i}$.

In the case of a catalogue which has been reduced to the HIPPARCOS system, the
errors $\varepsilon_{x, i}$ and $\varepsilon_{\mu, i}$ have to take into
account also the local uncertainty of the systematic corrections, e.g. by using
$\varepsilon^2_{x, i, tot} = \varepsilon^2_{x, i, ind} + \varepsilon^2_{x, i,
sys}$, where the `individual' error $\varepsilon_{x, i, ind}$ is the random
measuring error given in the catalogue. We should emphasize that
$\varepsilon_{x, i, sys}$ is not the systematic correction itself but only its
uncertainty.

The central epoch $T_i$ is chosen such that the correlation between $x_i (T_i)$
and $\mu_i$ is zero at $t = T_i$. Most of the ground-based catalogues give the
central epoch $T_i$ explicitely for each star, separately for $\alpha$ and
$\delta$. For HIPPARCOS (index 2) we have to derive
$T_2 = T_{H, ind}$ (separately for
$\alpha_\ast$ and $\delta$) from the data given in the catalogue:
\begin{equation}
T_2 = T_{H, ind} = T_H - \frac{\rho_{x\mu, H}\,(T_H)\,\varepsilon_{x,
H}\,(T_H)}{\varepsilon_{\mu, H}\,(T_H)} \,\, ,                          %(1)
\end{equation}
where $T_H = 1991.25$ is the overall reference epoch of the HIPPARCOS
Catalogue, $\varepsilon_{x, H}\,(T_H)$ the mean error of $x_H\,(T_H)$,
$\varepsilon_{\mu, H}$ the mean error of $\mu_H$, and $\rho_{x\mu, H}\,(T_H)$
the correlation coefficient between $x_H (T_H)$ and $\mu_H$. The position $x_2
(T_2)$ is derived from
\begin{equation}
x_2 (T_2) = x_H (T_H) + \mu_H (T_2 - T_H) \,\, .                        %(2)
\end{equation}
The mean error of $x_2 (T_2)$ is given by
\begin{eqnarray}
\varepsilon^2_{x, 2} (T_2)  & = &  \varepsilon^2_{x, H} (T_H) +
\varepsilon^2_{\mu, H} (T_2 - T_H)^2\nonumber\\
& + & 2 \, \rho_{x\mu, H} (T_H) \, \varepsilon_{x, H} (T_H) \,
\varepsilon_{\mu, H} (T_2 - T_H) \,\, .                                 %(3)
\end{eqnarray}
The proper motion has not to be recalculated, i.e. $\mu_2 = \mu_H$ and
$\varepsilon_{\mu, 2} = \varepsilon_{\mu, H}$. The HIPPARCOS position and
proper motion, $x_2 (T_2)$ and $\mu_2$, are now uncorrelated, i.e.
$\rho_{x\mu, 2} (T_2) = 0$. In the analytic approach, we assume that the
values of $x_2 (T_2)$ and $\mu_2$ between $\alpha_\ast$ and $\delta$ are also
uncorrelated. This is not strictly true because our choice of $T_2$ (separately
for $\alpha_\ast$ and $\delta$) forces only one correlation coefficient to
vanish. All the other correlation coefficients of the HIPPARCOS data remain
finite (albeit small) in general. Practical experience shows that this
approximation is quite good for most stars (see Sect. 5).

We are now prepared to carry out the combination of the two catalogues. The
basic idea of the combination is to reconstruct the normal equations of the
least-square solutions from which the two catalogues have been derived. This
can be done by using only the data actually given in the two catalogues. These
two sets of normal equations are then added together to provide the normal
equations for the combined catalogue C. Our method is a generalisation of
methods described by Kopff et al. (1964), Eichhorn (1974), and other authors.
First partial results were presented by Wielen (1988).

The method of least squares determines the `best' solution $x_{sol} (T_{ref})$
and $\mu_{sol}$ for the position of a star at a reference epoch $T_{ref}$ and
for its proper motion from a time series of observed positions,
\begin{equation}
x_{obs} (t) = x_{sol} (T_{ref}) + \mu_{sol} (t - T_{ref}) + v_{obs} (t) \,\, ,
\end{equation}                                                           %(4)
by the condition that the weighted sum of the squared residuals $v_{obs} (t)$
should be a minimum:
\begin{equation}
\left[p\,v^2_{obs}\right] = min \,\, .                                   %(5)
\end{equation}
We use here the classical convention that the brackets [...] imply a summation
over the observations. $p$ is the individual weight of $x_{obs} (t)$. (We
hope that the reader is not confused by the fact that we use the same letter
$p$ for two different quantities in order not to change familiar notations: In
this Sect. 2, we denote by $p$ the weights of observations. In all the other
sections, $p$ is the stellar parallax.) The condition (5) is fulfilled if we
solve the normal equations
for $x_{sol} (T_{ref})$ and $\mu_{sol}$:
\begin{eqnarray}
\left[p\right]\,x_{sol} (T_{ref})
& + & \,\,\,\, \left[p (t - T_{ref})\right]\,\mu_{sol}   \nonumber\\
&   & \hspace{0.60 cm} = \left[p\,x_{obs} (t)\right]    \,\, ,\\
                                                                        %(6)
\left[p (t - T_{ref})\right]\,x_{sol} (T_{ref})
& + & \left[p (t - T_{ref})^2\right]\,\mu_{sol} \nonumber\\
&   & \hspace{0.60 cm} = \left[p\,x_{obs} (t) (t - T_{ref})\right]  \, . %(7)
\end{eqnarray}
We now specialize the Eqs. (6) and (7) for each of the basic catalogues
(i\,=\,1, 2). For this purpose, we choose individual reference epochs $T_{ref}$
for each catalogue $(T_{ref} = T_i)$ such that
\begin{equation}
\left[p (t - T_i)\right]_i = 0 \,\, .                                   %(8)
\end{equation}
The index $i$ at the right bracket indicates that the sum should be
extended over all the observations which have been used in catalogue $i$. The
normal equations of catalogue $i$ are then
\begin{eqnarray}
\left[p\right]_i \, x_i (T_i) & = & \left[p\,x_{obs} (t)\right]_i \,\, \\ %(9)
\left[p (t - T_i)^2\right]\,\mu_i & = & \left[p\,x_{obs} (t) (t - T_i)\right]_i
\,\, .                                                                  %(10)
\end{eqnarray}
The choice (8) of the central epoch $T_i$ for $T_{ref}$ has the consequence
that $x_i (T_i)$ and $\mu_i$ are not correlated, as indicated by the vanishing
diagonal terms which leads to decoupled equations for $x_i (T_i)$ and $\mu_i$.

The mean errors $\varepsilon_{x, i}$ of $x_i (T_i)$ and $\varepsilon_{\mu, i}$
of $\mu_i$ are given by:
\begin{eqnarray}
\varepsilon^2_{x, i} & = & \frac{\varepsilon^2_0}{\left[p\right]_i} \,\, ,\\%11
\varepsilon^2_{\mu, i} & = & \frac{\varepsilon^2_0}{\left[p (t -
T_i)^2\right]_i} \,\, ,                                                 %(12)
\end{eqnarray}
where $\varepsilon_0$ is the error of unit weight. It will turn out that we do
not have to know $\varepsilon_0$ or the adopted system of weights in each of
the basic catalogues. We rely only on the fact that mean errors were correctly
calculated for the final results given in each of the catalogues.

The four quantities in the brackets of Eqs. (9) and (10) are now redetermined
in the following. From Eqs. (11) and (12), we obtain
\begin{eqnarray}
\left[p\right]_i
       & = & \frac{\varepsilon^2_0}{\varepsilon^2_{x, i}} \,\, , \\  %(13)
\left[p (t - T_i)^2\right]_i
       & = & \frac{\varepsilon^2_0}{\varepsilon^2_{\mu, i}} \,\, .   %(14)
\end{eqnarray}
Since the catalogue values of $x_i (T_i)$ and $\mu_i$ solve the Eqs. (9) and
(10), we can reconstruct the right-hand sides of the normal equations with the
help of Eqs. (13) and (14):
\begin{eqnarray}
\left[p\,x_{obs} (t)\right]_i
        & = & \frac{\varepsilon^2_0}{\varepsilon^2_{x, i}} \,
            x_i (T_i) \,\, , \\                                         %(15)
\left[p\,x_{obs} (t) (t - T_i)\right]_i
        & = & \frac{\varepsilon^2_0}{\varepsilon^2_
            {\mu, i}} \, \mu_i \,\, .                                   %(16)
\end{eqnarray}

The normal equations for the combined catalogue C have the same form as Eqs.
(6) and (7). In order to simplify these normal equations, we replace $T_{ref}$
by the central epoch $T_C$ of the combined catalogue C. $T_C$ is given by the
condition:
\begin{equation}
\left[p (t - T_C)\right]_C = 0 \,\, .                                   %(17)
\end{equation}
$T_C$ can be determined from known quantities by using Eqs. (8):
\begin{eqnarray}
\lefteqn{
       \left[p (t - T_C)\right]_C
    =  \left[p (t - T_C)\right]_1
    +  \left[p (t - T_C)\right]_2
}
                                   \nonumber\\
& & =  \left[p (t - T_1)\right]_1
    +  \left[p (T_1 - T_C)\right]_1\nonumber\\
& & +  \left[p (t - T_2)\right]_2
    +  \left[p (T_2 - T_C)\right]_2\nonumber\\
& & =  \left[p\right]_1 (T_1 - T_C)
    +  \left[p\right]_2 (T_2 - T_C) = 0 \,\, .
\end{eqnarray}                                                          %(18)
Solving for $T_C$ and inserting Eqs. (13), we find
\begin{eqnarray}
T_C & = &  \frac{\left[p\right]_1\,T_1 + \left[p\right]_2\,T_2}
               {\left[p\right]_1 + \left[p\right]_2}
      =  \frac{{\varepsilon^{-2}_{x, 1}} \, T_1 +
                {\varepsilon^{-2}_{x, 2}} \, T_2}
                {{\varepsilon^{-2}_{x, 1}} + {\varepsilon^{-2}_{x,2}}}
      \nonumber \\
    & = & \frac{w_{x, 1}\, T_1 + w_{x, 2}\, T_2}{w_{x, 1} +
                 w_{x,2}} \,\, ,                                        %(19)
\end{eqnarray}
The central epoch $T_C$ is therefore the weighted mean of the epochs $T_1$ and
$T_2$. As weights for $T_i$, we have to use the weight
\begin{equation}
w_{x, i} = \frac{1}{\varepsilon^2_{x, 1}}                               %(20)
\end{equation}
of the position $x_i (T_i)$. Using $T_C$ as the reference time, the normal
equations for the combined catalogue C are reduced to two independent equations
for the unknowns $x_C (T_C)$ and $\mu_C$, the position and proper motion of the
star in C:
\begin{eqnarray}                                                        %(21)
\left[p\right]_C \, x_C (T_C) & = & \left[p \, x_{obs} (t)\right]_C \,\, ,\\
\left[p (t - T_C)^2\right]_C \, \mu_C & = & \left[p \, x_{obs} (t) (t
- T_C)\right]_C \,\, .                                                  %(22)
\end{eqnarray}
If we solve Eq. (21) for $x_C (T_C)$ and use Eqs. (9), (13), and (15), we find
\begin{eqnarray}
x_C (T_C) & = & \frac{\left[p\,x_{obs}
(t)\right]_C}{\left[p\right]_C}
            =   \frac{\left[p\,x_{obs} (t)\right]_1 + \left[p\,x_{obs}
(t)\right]_2}{\left[p\right]_1 + \left[p\right]_2} \nonumber \\
          & = & \frac{\left[p\right]_1\,x_1 (T_1) + \left[p\right]_2\,x_2
(T_2)}{\left[p\right]_1 + \left[p\right]_2} \nonumber \\
          & = & \frac{w_{x, 1}\,x_1 (T_1) + w_{x, 2}\,x_2 (T_2)}{w_{x, 1} +
w_{x,2}} \,\, ,                                                         %(23)
\end{eqnarray}
with $w_{x, i}$ given by Eq. (20). As for $T_C$, the position $x_C (T_C)$ is
the weighted mean of $x_1 (T_1)$ and $x_2 (T_2)$ with $w_{x, i}$ as weights.

In order to write $\mu_C$ also as a weighted average, it is helpful to
introduce a third proper motion, $\mu_0$, as an auxiliary tool (Wielen 1988):
\begin{equation}
\mu_0 = \frac{x_2 (T_2) - x_1 (T_1)}{T_2 - T_1} \,\, .                  %(24)
\end{equation}
The proper motion $\mu_0$ is based only on the central positions of the two
basic catalogues (see also Fig. 1).
Since $x_1 (T_1)$ and $x_2 (T_2)$ are not correlated with the
catalogue proper motions $\mu_1$ and $\mu_2$, the new proper motion $\mu_0$ is
also not correlated with $\mu_1$ and $\mu_2$. The mean error
$\varepsilon_{\mu, 0}$ of $\mu_0$ is given by
\begin{equation}
\varepsilon^2_{\mu, 0} = \frac{\varepsilon^2_{x, 1} + \varepsilon ^2_{x, 2}}
{(T_2 - T_1)^2} \,\, .                                                  %(25)
\end{equation}
The weights $w_{\mu, i}$ of the proper motions are
\begin{equation}
w_{\mu, i} = \frac{1}{\varepsilon^2_{\mu, i}} \,\, ,                    %(26)
\end{equation}
where the index $i$ now runs from 0 to 2. We now solve Eq. (22) for $\mu_C$:
\begin{equation}
\mu_C = \frac{\left[p\,x_{obs} (t) (t - T_C)\right]_C}{\left[p (t -
T_C)^2\right]_C} \,\, .                                                 %(27)
\end{equation}
For the numerator of Eq.(27)
we obtain by using Eqs. (15), (16), (19), (20), (24),
(25), and (26):
\begin{eqnarray}
\lefteqn{
\left[p\,x_{obs} (t) (t - T_C)\right]_C
}
\nonumber\\
& & = \left[p\,x_{obs} (t) (t - T_C)\right]_1 + \left[p\,x_{obs} (t) (t -
T_C)\right]_2 \nonumber\\
& & = \left[p\,x_{obs} (t) (t - T_1)\right]_1 + \left[p\,x_{obs} (t) (t -
T_2)\right]_2\nonumber\\
& & \hspace{1.00cm}
+ \,(T_1 -T_C) \left[p\,x_{obs} (t)\right]_1 + (T_2 - T_C) \left[p\,x_{obs}
(t)\right]_2 \nonumber\\[1ex]
& & = \varepsilon ^2_0 \, \Big(w_{\mu,1}\,\mu_1 + w_{\mu, 2}\,\mu_2
+ (T_1 - T_C)\,w_{x,1}\,x_1 (T_1)\nonumber\\
& & \hspace{1.00cm}
+\, (T_2 - T_C)\,w_{x, 2} \, x_2 (T_2)\Big) \nonumber\\[1ex]
& & = \varepsilon^2_0 \, \Big(w_{\mu, 1} \,
\mu_1 + w_{\mu, 2} \, \mu_2\nonumber\\
& & \hspace{1.00cm}
+ \, \frac{w_{x, 2}}{w_{x, 1}
+ w_{x, 2}} \, (T_1 - T_2) \, w_{x, 1} \, x_1 (T_1)  \nonumber\\
& & \hspace{1.00cm}
+ \, \frac{w_{x, 1}}{w_{x, 1}
+ w_{x, 2}} \, (T_2 - T_1) \, w_{x, 2} \, x_2 (T_2)\Big) \nonumber\\[1ex]
& & = \varepsilon^2_0 \, \Big(w_{\mu, 1} \, \mu_1
+ w_{\mu, 2} \, \mu_2  \nonumber\\
& & \hspace{1.00cm}
+ \, \frac{w_{x, 1}\, w_{x, 2} \, (T_2 - T_1)^2}{w_{x, 1}
+ w_{x, 2}} \,\,\,
\frac{x_2 (T_2) - x_1 (T_1)}{T_2 - T_1}\Big)  \nonumber\\
& & = \varepsilon^2_0 \, \Big(w_{\mu, 1} \, \mu_1 + w_{\mu, 2} \, \mu_2
+ \, \frac{(T_2 - T_1)^2}{\varepsilon^2_{x, 1} + \varepsilon^2_{x, 2}} \,
\mu_0\Big)  \nonumber\\
& & = \varepsilon^2_0 \, (w_{\mu, 1} \, \mu_1 + w_{\mu, 2} \, \mu_2
+ w_{\mu, 0} \, \mu_0) \,\, .                                           %(28)
\end{eqnarray}
Similar to Eq. (28) we obtain for the denominator of Eq. (27) after some
algebra
\begin{eqnarray}
\lefteqn{
\left[p (t - T_C)^2\right]_C = \left[p (t - T_C)^2\right]_1
+ \left[p (t - T_C)^2\right]_2
}
\nonumber\\
& & = \left[p \big((t - T_1) + (T_1 - T_C)\big)^2\right]_1  \nonumber\\
& & \hspace{1.00cm}
    + \left[p \big((t - T_2) + (T_2 - T_C)\big)^2\right]_2  \nonumber\\
& & = \left[p (t - T_1)^2\right]_1 + \left[p (t - T_2)^2\right]_2\nonumber\\
& & \hspace{1.00cm}
    + \, (T_1 - T_C)^2 \, \left[p\right]_1 + (T_2 -T_C)^2 \,
      \left[p\right]_2                              \nonumber\\
& & = \varepsilon^2_0 \,
(w_{\mu, 1} + w_{\mu, 2} + w_{\mu, 0}) \, . \,\,                         %(29)
\end{eqnarray}
Our final result for $\mu_C$ is therefore, using Eqs. (27)\,-\,(29),
\begin{equation}
\mu_C = \frac{w_{\mu, 1} \, \mu_1 + w_{\mu, 2} \, \mu_2 + w_{\mu, 0} \,
\mu_0}
{w_{\mu, 1} + w_{\mu, 2} + w_{\mu, 0}} \,\, .                           %(30)
\end{equation}
Equation (30) means that the combined proper motion $\mu_C$ is the weighted
average of the three proper motions $\mu_1, \mu_2, \mu_0$, where we have to use
Eq. (26) for the weights of the proper motions.

The mean errors $\varepsilon_{x, C}$ and $\varepsilon_{\mu, C}$ of the combined
position $x_C (T_C)$ and combined
proper motion $\mu_C$ are derived from Eqs. (11),
(12), and (29):
\begin{eqnarray}
\varepsilon^2_{x, C}   & = &  \frac{\varepsilon^2_0}{\left[p\right]_C} =
\frac{\varepsilon^2_0}{\left[p\right]_1 + \left[p_2\right]}
                       =   \frac{1}{{\varepsilon^{-2}_{x, 1}} +
                           {\varepsilon^{-2}_{x, 2}}} \,\, ,\\          %(31)
\varepsilon^2_{\mu, C} & = &  \frac{\varepsilon^2_0}{\left[p (t -
                                          T_C)^2\right]_C}
                                                            \nonumber\\
                       & = & \frac{1}{w_{\mu, 1} + w_{\mu, 2} + w_{\mu, 0}}
                         =   \frac{1}{{\varepsilon^{-2}_{\mu, 1}} +
{\varepsilon^{-2}_{\mu, 2}} + {\varepsilon^{-2}_{\mu, 0}}} \,\, .       %(32)
\end{eqnarray}
The weights of $x_C (T_C)$ and $\mu_C$ are therefore just the sum of the
weights of the two positions,
\begin{equation}
w_{x, C} = w_{x, 1} + w_{x, 2} \,\, ,                                   %(33)
\end{equation}
and the sum of the weights of the three proper motions,
\begin{equation}
w_{\mu, C} = w_{\mu, 1} + w_{\mu, 2} + w_{\mu, 0} \,\, .                %(34)
\end{equation}
Due to the separation of the normal equations (21) and (22) by our choice of
$T_C$, the values of $x_C (T_C)$ and $\mu_C$ are not correlated.

In summary, our analytic approach to the combination of the two basic
catalogues provides a very simple rule: the resulting combined data are
weighted averages of the data in the two catalogues, described by Eqs. (19),
(23), and (30). For the proper motions we have to include the proper motion
$\mu_0$, based on the two positions, into this average. The mean errors of the
combined position and proper motion are given by Eqs. (31) and (32).
A schematic illustration of the method is shown in Fig. 1.
While our
method is pleasing because of its conceptional
simplicity, it is also fully justified by the use of the normal equations. As
mentioned earlier, the only approximation used in our analytic approach is that
we have neglected the cross-correlations
between the coordinate components of the
positions
and proper motions in $\alpha_\ast$ and $\delta$ introduced by HIPPARCOS.

As a preparation for Sect. 3, we remark
(without giving the proof here)
that our analytic result is
identical to the solution of another least-square problem in which we treat the
catalogue data $x_1 (T_1), \, \mu_1, \, x_2 (T_2), \, \mu_2$ as `observations'.
The new condition is
\begin{eqnarray}
w_{x, 1} \, \big(x_1 (T_1) - x_C (T_1)\big)^2 + w_{x, 2} \big(x_2 (T_2) - x_C
(T_2)\big)^2\nonumber\\
+  w_{\mu, 1} \, (\mu_1 - \mu_C)^2 + w_{\mu, 2} \, (\mu_2 - \mu_C)^2 = min
\,\, .                                                                  %(35)
\end{eqnarray}
The weights $w$ are given by Eqs. (20) and (26), and $x_C (T_i)$ follows from
\begin{equation}
x_C (T_i) = x_C (T_C) + \mu_C (T_i - T_C) \,\, .                        %(36)
\end{equation}
Finally we would like to point out that our analytic method can be easily used
for a `decomposition' of a combined catalogue: If the catalogues C and 1 are
available, then
the catalogue 2 can be fully reconstructed. In subsequent papers we
shall present results of this `inverse combination' method by decomposing
earlier fundamental catalogues. For example, we have decomposed the FK5 into
two parts: the published FK4 and the `summary of subsequent data' which we
denote by FK5-4. The FK5-4 has also the character of a compilation catalogue.
The decomposition of the FK5 into a few `subcatalogues' has the
following purpose: If we would like to test the motion of a fundamental star
for linearity in time or to derive a non-linear motion for the star similar to
the G solutions of HIPPARCOS, then it is helpful to have as many catalogues
with different, well-spaced epochs as possible at hand.

%%%%%%%%%%%%%%%%%%%%%%%%%%%%%%%%% Begin of Figure 1 %%%%%%%%%%%%%%%%%%%%%%
% Begin of        %
%                 %
% F i g u r e   1 %
%                 %
%%%%%%%%%%%%%%%%%%%
\begin{figure}[t]
\resizebox{\hsize}{!}{\includegraphics{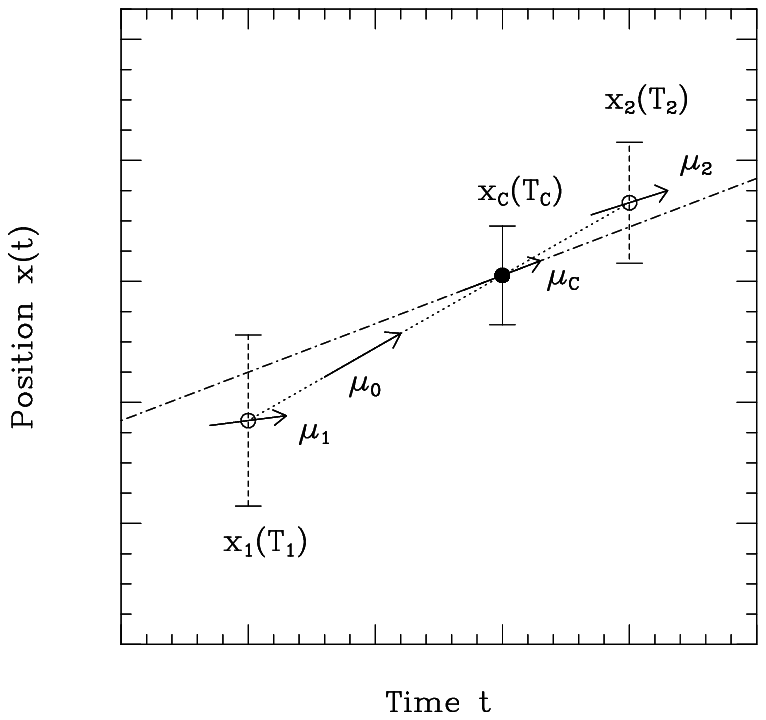}}
\caption[]
{Schematic illustration of the analytic approach to the combination of two
catalogues. The basic data are the positions of a star, $x_1$ and $x_2$ at the
two epochs $T_1$ and $T_2$, and the corresponding proper motions $\mu_1$ and
$\mu_2$. The dotted line connects the two positions (open circles). The
combined position $x_C$ (filled circle) at the combined central epoch $T_C$ is
the weighted average of $x_1$ and $x_2$ and is located therefore on this
connecting line. For the positions, error bars are shown. The dotted line
provides also the direction of the additional proper motion $\mu_0$, derived
from the two positions. The combined proper motion $\mu_C$ is the weighted
average of the three proper motions $\mu_1,\,\mu_2,\,\mu_0$. The dash-dotted
line illustrates the combined solution $x_C (t)$ as a function of time t.}
\end{figure}
%%%%%%%%%%%%%%%%%%%%%%%%%%%%%%%% End of Figure 1 %%%%%%%%%%%%%%%%%%%%%%%%%%%%%

%Section 3
\section{Numerical approach}

We now present our numerical approach to the problem of combining two
catalogues. In this numerical approach, the correlations between the five
astrometric parameters derived by HIPPARCOS in a standard solution for an
apparently single star are fully taken into account.

We change our notation slightly. As a ground-based catalogue, we envisage the
FK5, and use the index $F$ (instead of 1). The HIPPARCOS data receive the index
$H$ (instead of 2). The index $C$ for the combined catalogue remains.
Furthermore we will use now the vector and matrix notation in most cases. {\bf
U} is the unit matrix with the elements $U_{kl} = \delta_{kl}\,(k, l =
1, 2, 3, ...)$; {\bf 0} is
a submatrix filled with zeros. ${\bf A}^T$ denotes the transpose of the matrix
${\bf A}$, and ${\bf A}^{-1}$ its inverse, ${\bf AA}^{-1} = {\bf U}$. Instead
of the right ascension
$\alpha$ and the corresponding proper motion $\mu_{\alpha}$, we use
$\alpha_{\ast} = \alpha\,\cos\,\delta$ and $\mu_{\alpha\ast} =
\mu_{\alpha}\,\cos\,\delta$. The parallax of a star is denoted by $p$.

As already mentioned in Sect. 1, we use in real applications only quantities
which are differential with respect to HIPPARCOS, e.g. $\alpha_{\ast, F}
(t) - \alpha_{\ast, H} (t)$ instead of $\alpha_{\ast, F} (t)$. This means, of
course, that most of the quantities with the index $H$ are zero in our scheme.
Here we keep, however, the full quantities in order not to confuse the reader.

The HIPPARCOS Catalogue provides the 5 astrometric parameters
\begin{eqnarray}
{\bf b}_H \, (T_H) = \left( \begin{array}{c}
                     \alpha_{\ast, H} (T_H)\\
                     \delta_H (T_H)\\
                     \mu_{\alpha\ast, H}\\
                     \mu_{\delta, H}\\
                     p_H
                      \end{array}   \right)                             %(37)
\end{eqnarray}
for the overall epoch $T_H = 1991.25$. Furthermore, the HIPPARCOS Catalogue
gives all the data required to derive the corresponding variance-covariance
matrix ${\bf D}_H (T_H)$. The diagonal line of ${\bf D}_H$ is occupied by the
squares of the mean errors of the components of ${\bf b}_H (T_H)$, e.g. by
\begin{equation}
D_{H, 11} = \varepsilon^2_{\alpha\ast, H} \, (T_H) \,\, .               %(38)
\end{equation}
The non-diagonal elements of ${\bf D}_H$ can be derived from the correlation
coefficients $\rho_{kl, H} \, (k, l = 1,..., 5)$ and the mean errors, e.g. by
\begin{equation}
D_{H, 12} = D_{H, 21} = \varepsilon_{\alpha\ast, H} (T_H) \,
\varepsilon_{\delta, H} (T_H) \, \rho_{\alpha\delta, H} (T_H) \,\, .
%(39)
\end{equation}
The FK5 provides only 4 astrometric parameters,
\begin{eqnarray}
{\bf b}_F = \left( \begin{array}{c}
                   \alpha_{\ast, F} (T_{\alpha, F})\\
                   \delta_F (T_{\delta, F})\\
                   \mu_{\alpha\ast, F}\\
                   \mu_{\delta, F} \end{array} \right) \,\, ,           %(40)
\end{eqnarray}
since no parallax is given. In contrast to HIPPARCOS, the positions are given
at the central epochs, $T_{\alpha, F}$ and $T_{\delta, F}$. Hence the four FK5
astrometric parameters are not correlated. This means that all the non-diagonal
elements of the variance-covariance matrix ${\bf D}_F$ are zero. The diagonal
elements of ${\bf D}_F$ are given by the squares of the mean errors, e.g. by
\begin{equation}
D_{F, 11} = \varepsilon^2_{\alpha\ast, F} \, (T_{\alpha, F}) \,\, .     %(41)
\end{equation}
The total vector {\bf b} of `observations' (9 elements) is given by
\begin{eqnarray}
{\bf b} = \left( \begin{array}{l}
                 {\bf b}_H\\
                 {\bf b}_F \end{array} \right) \,\, ,                   %(42)
\end{eqnarray}
and the total matrix ${\bf D}$ (9\,$\times$\,9 elements) by
\begin{eqnarray}
{\bf D} = \left( \begin{array}{ll}
                 {\bf D}_H & {\bf O}\\
                 {\bf O}   & {\bf D}_F \end{array} \right) \,\, ,       %(43)
\end{eqnarray}
since the HIPPARCOS and FK5 data have no cross-cor\-re\-la\-tions.
The total matrix
${\bf P}$ of the weights is the inverse of ${\bf D}$:
\begin{equation}
{\bf P} = {\bf D}^{-1} \,\, .                                           %(44)
\end{equation}
As the unknowns to be determined, we use the two positions at an arbitrary
reference epoch, $T_{ref}$, the two proper motions, and the parallax:
\begin{eqnarray}
{\bf c} = \left( \begin{array}{c}
                 \alpha_{\ast, C} (T_{ref})\\
                 \delta_C (T_{ref})\\
                 \mu_{\alpha\ast, C}\\
                 \mu_{\delta, C}\\
                 p_C \end{array} \right) \,\, .                         %(45)
\end{eqnarray}
We now write down the equations of conditions for each `observation',
similar to Eq. (4) of our analytic approach. In matrix form the equations of
conditions are
\begin{equation}
{\bf b} = {\bf A} \, {\bf c} + {\bf v} \,\, .                           %(46)
\end{equation}
The design matrix {\bf A} is given by
\begin{eqnarray}
{\bf A} = \left( \begin{array}{ccccc}
                 1 \hspace*{0.5cm} & 0 & \hspace*{0.2cm}T_H -
                                            T_{ref} & 0 & \hspace*{0.2cm} 0\\
                 0 \hspace*{0.5cm} & 1 & \hspace*{0.2cm}0 & T_H -
                                                T_{ref} & \hspace*{0.2cm} 0\\
                 0 \hspace*{0.5cm} & 0 & \hspace*{0.2cm} 1 & 0 &
                                                          \hspace*{0.2cm} 0\\
                 0 \hspace*{0.5cm} & 0 & \hspace*{0.2cm} 0 & 1 &
                                                          \hspace*{0.2cm} 0\\
                 0 \hspace*{0.5cm} & 0 & \hspace*{0.2cm} 0 & 0 &
                                                          \hspace*{0.2cm} 1\\
                 1 \hspace*{0.5cm} & 0 & \hspace*{0.2cm} T_{\alpha, F} -
                                             T_{ref} & 0 & \hspace*{0.2cm} 0\\
                 0 \hspace*{0.5cm} & 1 & \hspace*{0.2cm} 0 & T_{\delta, F} -
                                                 T_{ref} & \hspace*{0.2cm} 0\\
                 0 \hspace*{0.5cm} & 0 & \hspace*{0.2cm} 1 & 0 &
                                                           \hspace*{0.2cm} 0\\
                 0 \hspace*{0.5cm} & 0 & \hspace*{0.2cm} 0 & 1 &
                            \hspace*{0.2cm} 0 \end{array} \right) \,\, .  %(47)
\end{eqnarray}
The vector {\bf v} contains the residuals of the 9 observational parameters.
The least-square solution of Eq. (46) requires that the sum of the weighted
squares of the residuals is minimized:
\begin{equation}
{\bf v}^T \, {\bf P \, v} = min \,\, .                                  %(48)
\end{equation}
From this condition the normal equations follow:
\begin{equation}
{\bf A}^T \, {\bf P} \, {\bf A} \, {\bf c} = {\bf A}^T \, {\bf P} \, {\bf b}
\,\, .                                                                  %(49)
\end{equation}
The solution {\bf c} of this equation gives the five astrometric parameters of
the combined catalogue C for the epoch $T_{ref}$. The mean errors and the
correlation coefficients of the elements of {\bf c} are derived from the
variance-covariance matrix of {\bf c} which is given by
\begin{equation}
{\bf D}_C = ({\bf A}^T \, {\bf P} \, {\bf A})^{-1} \,\, .               %(50)
\end{equation}
For example,
\begin{equation}
\varepsilon^2_{\alpha\ast, C} (T_{ref}) = D_{C, 11} \,\, ,              %(51)
\end{equation}
or
\begin{equation}
\rho_{\alpha\delta, C} (T_{ref}) = D_{C, 12}\,/\,(D_{C, 11} \, D_{C, 22})^{1/2}
\,\, .                                                                  %(52)
\end{equation}
The positions, their mean errors and their correlation coefficients can, of
course, be calculated for other epochs than $T_{ref}$, say for an epoch $T$, by
using
\begin{equation}
\alpha_{\ast, C} (T) = \alpha_{\ast, C} (T_{ref}) + \mu_{\alpha\ast, C} (T -
T_{ref})                                                                %(53)
\end{equation}
and a corresponding equation for $\delta_C (T)$. A specially interesting epoch
is $T_{\alpha, C, min}$ at which the mean error
$\varepsilon_{\alpha\ast, C}$ of $\alpha_{\ast, C}$ reaches its minimum. This
epoch corresponds to $T_C$ for $\alpha_\ast$ in our analytic approach. In
general, $T_{\alpha, C, min}$ and $T_{\delta, C, min}$ are not the same
epochs.

The numerical approach is well adapted for many variations. Firstly, we can
easily add further compilation catalogues or individual observational
catalogues. Secondly, we can remove unknowns or add further unknowns. For
example, the `long-term prediction mode' of the FK6 requires that we drop the
parallax $p$ as an unknown. This can be done either explicitly in our
equations or by giving $p_H$ a weight zero, corresponding formally to
$\varepsilon_{p, H}
\rightarrow \infty$. Additional unknowns, e.g. acceleration
terms similar to the $G$ solutions of HIPPARCOS, are required if we would like
to determine non-linear solutions for some astrometric binaries. The use of the
`decomposed' FK5 mentioned at the end of Sect. 2, demands an increase in both
the number of catalogues and the number of unknowns.

%%%%%%%%%%%%%%%%%%%
% Begin of        %
%                 %
% T a b l e  1    %
%                 %
%%%%%%%%%%%%%%%%%%%

\begin{table*}[th]
\caption
{
Results of the combination method for the star HIP 9884 = FK 74 = GC 2538 =
$\alpha$ Ari
}
\begin{tabular}{lrrrrrrrrrrrr}
\hline\\[-1.0ex]
& \multicolumn{6}{c}{FK5 + HIPPARCOS} & \multicolumn{6}{c}{GC + HIPPARCOS}\\
Quantity & $\alpha_\ast$ & \multicolumn{2}{c}{mean error} &
$\delta$ & \multicolumn{2}{c}{mean error} & $\alpha_\ast$ &
\multicolumn{2}{c}{mean error} & $\delta$ & \multicolumn{2}{c}{mean error}
\\[0.5ex]\hline\\[-1.8ex] Input data:\\[1ex]
$\Delta x_F (T_F)$ or $\Delta x_{GC} (T_{GC})$ & -- 7.83 & $\pm$ 12.51 && +
96.94 & $\pm$ 14.78 && -- 75.60 & $\pm$ 39.62 && + 237.68 & $\pm$ 27.01 &\\
$\Delta\mu_F$ or $\Delta\mu_{GC}$ & + 0.49 & $\pm$\hphantom{1} 0.40 &&
--\hphantom{1} 1.20 & $\pm$\hphantom{1} 0.38 && +\hphantom{7} 1.70 &
$\pm$\hphantom{3} 0.99 && --\hphantom{23} 2.83 & $\pm$\hphantom{2} 0.95 &\\
$T_F$ or $T_{GC}$ & 1947.84 & & & 1929.73 & & & 1892.60 && & 1890.30
&&\\[1.5ex]
$\Delta x_H (T_{H, ind})$  & 0.00 & $\pm$ 0.77 && 0.00 & $\pm$ 0.54 && 0.00 &
$\pm$ 0.77 && 0.00 & $\pm$ 0.54 &\\
$\Delta\mu_H$ & 0.00 & $\pm$ 1.01 && 0.00 & $\pm$ 0.77 && 0.00 & $\pm$ 1.01 &&
0.00 & $\pm$ 0.77 &\\
$T_{H, ind}$ & 1991.26 & && 1991.51 && & 1991.26 && & 1991.51 &&\\[1.5ex]
$p_H$ & 49.48 & $\pm$ 0.99 &&& & & 49.48 & $\pm$ 0.99 &&& &\\[2ex]
Results from analytic approach:\\[1ex]
$\Delta\mu_0$ & + 0.18 & $\pm$ 0.29 && -- 1.57 & $\pm$ 0.25 && + 0.77 & $\pm$
0.40 && -- 2.35 & $\pm$ 0.27 &\\[1.5ex]
$\Delta x_C (T_C)$ & -- 0.03 & $\pm$ 0.77 && + 0.12 & $\pm$ 0.54 && -- 0.03 &
$\pm$ 0.77 && + 0.09 & $\pm$ 0.54 &\\
$\Delta\mu_C$ & + 0.27 & $\pm$ 0.23 && -- 1.36 & $\pm$ 0.20 && + 0.79 & $\pm$
0.35 && -- 2.14 & $\pm$ 0.24 &\\
$T_C$ & 1991.10 && & 1991.44 && & 1991.22 && & 1991.47 &&\\[2ex]
Results from numerical approach:\\[1ex]
$\Delta x_C (T_C)$ & -- 0.18 & $\pm$ 0.76 && + 0.14 & $\pm$ 0.54 && -- 0.27 &
$\pm$ 0.77 && + 0.15 & $\pm$ 0.54 &\\
$\Delta\mu_C$ & + 0.23 & $\pm$ 0.23 && -- 1.34 & $\pm$ 0.20 && + 0.66 & $\pm$
0.35 && -- 2.10 & $\pm$ 0.24 &\\
$T_C$ & 1991.12 && & 1991.47 && & 1991.25 && & 1991.50 &&\\[1.5ex]
$\Delta p_C$ & -- 0.57 & $\pm$ 0.95 & & &&& -- 0.92 & $\pm$ 0.95 &&&
&\\[0.5ex]\hline\\[-1ex]
\end{tabular}
Units: mas, mas/year, or years
\end{table*}

%%%%%%%%%%%%%%%%%%%%%%%%%%%%%% End of Table 1 %%%%%%%%%%%%%%%%%%%%%%%%%%%%%

%Section 4
\section{The effect of cosmic errors}

The combination methods presented in Sects. 2 and 3 are valid for single
stars only, which move linearily in time on straight lines in space. We
call this version of our combination method the `single-star mode'. Some
binaries can be handled without any significant change in this mode of our
combination method if we use appropriate `reference points' for the binary,
e.g. the center-of-mass of the double star. The specific treatment of known
binaries will be presented elsewhere.

As mentioned in Sect. 1, undetected binaries introduce cosmic errors into the
positions and proper motions, especially into the `instantaneous' HIPPARCOS
data. In a subsequent paper we shall discuss the modifications of our
combination method which are required if we take such cosmic errors into
account. In many applications, it is only necessary to replace the actual
measuring errors of some quantities by formal errors which include both the
measuring and the cosmic errors of these quantities.

%Section 5
\section{An example: \object{$\alpha$ Ari}}

In order to illustrate our combination method we give in
Table 1 the results for
one individual star. All the positions and proper motions are given
differentially to the HIPPARCOS results for the given epoch. Since this is only
a numerical convenience, the mean errors of the quantities given in
Table 1 are
those of the full quantities themselves. The mean errors of the ground-based
quantities, e.g. of $x_F,\,\mu_F,\,\mu_0$, are `total' values which contain
both the individual measuring errors and the uncertainty in the systematic
corrections. The ground-based data given are already reduced to the HIPPARCOS
system. Hence the differences in Table 1 contain only the `individual'
differences, not anymore a `systematic' part. In order to obtain the final
result of our combination method, one should add the results with the index $C$
to the full HIPPARCOS data. e.g. $x_C (T_C) = \Delta x_C (T_C) + x_H (T_C)$ or
$\mu_C = \Delta\mu_C + \mu_H$.

The comparison of the results of the analytic approach with those of the
numerical approach shows a very good agreement. The numerical approach has made
use of the correlation coefficients between the HIPPARCOS data for
$\alpha_{\ast}$(1991.25), $\delta$(1991.25),
$\mu_{\alpha\ast},\,\mu_{\delta},\,p$ (+0.20, $-$0.01, +0.10, +0.26,
+0.00, $-$0.35, $-$0.25, +0.33, +0.02, +0.29). The effect of
the correlations is therefore small. This holds for most stars. The correlation
coefficients for the combined data for $\alpha_{\ast}$(1991.12),
$\delta$(1991.47), $\mu_{\alpha\ast},\,\mu_{\delta},\,p$ (+0.26, 0.00, +0.04,
+0.24, +0.04, 0.00, $-$0.16, +0.02, $-$0.01, +0.08) are smaller than for
HIPPARCOS, because the ground-based data are not correlated at all if we use
the concept of central epochs.

A comparison of the mean errors of the combined quantities with those of
HIPPARCOS shows the following: (1) There is a significant gain in the accuracy
of the proper motion which justifies the combination of HIPPARCOS results with
ground-based data. In our example, the mean errors of the combined proper
motions are smaller than those of the HIPPARCOS proper motion by a factor of
1/4.4 in $\mu_{\alpha\ast}$ and of 1/3.9 in $\mu_{\delta}$. (2) Since the mean
measuring errors of $x_H$ are extremely small, there is practically no gain in
the accuracy of the
central positions, and we have $T_C \sim T_H$. However, for
predicting positions at epochs which differ from $T_H$ by more than a few
years, the accuracy of the proper motion is governing the error of these
predicted positions. (3) The improvement of the HIPPARCOS parallax $p_H$ is
here and in most other cases not very significant, since the correlation of
$p_H$ with the other HIPPARCOS data $(x_H,\,\mu_H)$ is usually small.

The comparison of the combinations FK5+HIPP and GC+HIPP shows the
importance of the old observations. The FK5 contains essentially all the
observations compiled in the GC, and in addition many, more accurate, modern
observations. Nevertheless, the mean errors of the combined data for the
GC+HIPP are only slightly larger than for FK5+HIPP. The analytic
approach shows the reason: The accuracy of the proper motion $\mu_0$ is already
very high for the GC+HIPP, mainly because of the large epoch difference
of the central positions.

%Section 6
\section{Error budget}

In order to show the overall improvement in the
accuracy of the proper motions obtained
by using our combination method, we present here the statistics of the
appropriate mean errors for a few samples of stars.

The mean errors $\varepsilon_{\mu}$ of the proper motions $\mu$, given in
Tables
2 and 3, refer to one `mean' coordinate component of $\mu$. For each star, the
`mean' value $\varepsilon_{\mu, 1 {\rm D}}$ is obtained as a root-mean-square
(rms) average over $\varepsilon_{\mu, \alpha\ast}$ and $\varepsilon_{\mu,
\delta}$. Then these individual values of $\varepsilon_{\mu, 1 {\rm D}}$ are
averaged over the sample of stars under consideration, either by taking an rms
average over all the $\varepsilon_{\mu, 1 {\rm D}}$ or by selecting the median
value of $\varepsilon_{\mu, 1 {\rm D}}$ in the sample.

In Table 2, we present the error budget for the combination of the FK5 with
HIPPARCOS for the basic FK5 stars. This corresponds to the results of the
`single-star mode' of the FK6 (Wielen et al. 1998). We consider two samples of
stars: (a) all the 1535 basic FK5 stars, and (b) 1202 `apparently single' basic
FK5 stars. The error budget of all the basic FK5 stars is slightly fictious in
so far as this sample also contains binaries which cannot be safely treated by
a direct combination method. Nevertheless, the error budget for the 1535 stars
is still a valid indicator for the overall
accuracy of our combination method. The
error budget for the 1202 stars is slightly biased in favour of the HIPPARCOS
Catalogue, since most of the stars for which especially HIPPARCOS is less
accurate than usually (e.g., C solutions for visual binaries) are removed
from this sample. Table 2 shows that the FK6 proper motions, obtained from the
combination of the FK5 data with the HIPPARCOS observations, are typically
by a factor of about two more accurate than the proper motions in either the
FK5 alone or in HIPPARCOS Catalogue alone. Already $\mu_0$ is typically
slightly more accurate than $\mu_H$.

%%%%%%%%%%%%%%%%%%%
% Begin of        %
%                 %
% T a b l e    2  %
%                 %
%%%%%%%%%%%%%%%%%%%
\begin{table}[th]
\caption{
Error budget for FK6 proper motions
in the `single-star mode'
}
\begin{tabular}{lcccc}
\hline\\[-1.0ex]
\multicolumn{5}{c}{Typical mean errors of proper motions}\\
\multicolumn{5}{c}{(in one component, averaged over $\mu_{\alpha\ast}$ and
$\mu_\delta$; units: mas/year)}\\[1.5ex]\hline\\[-2.0ex]
Sample of stars:& \multicolumn{2}{c}{1535 FK} & \multicolumn{2}
{c}{1202 FK}\\ [0.7ex]
\hline\\[-0.5ex]
& rms aver. & median & rms aver. & median\\[0.5ex]\hline\\[-0.5ex]
HIPPARCOS& 0.82  & 0.63 & 0.68 & 0.61\\[1.5ex]
FK5\\
\hspace*{0.5cm}random& 0.76 & 0.64 & 0.77 & 0.67\\
\hspace*{0.5cm}system& 0.28 & 0.25 & 0.28 & 0.25\\
\hspace*{0.5cm}total & 0.81 & 0.70 & 0.83 & 0.72\\[1.5ex]
$\mu_0$\\
\hspace*{0.5cm}random& 0.53 & 0.43 & 0.54 & 0.45\\
\hspace*{0.5cm}system& 0.24 & 0.23 & 0.25 & 0.23\\
\hspace*{0.5cm}total & 0.58 & 0.49 & 0.59 & 0.51\\[4.0ex]
FK6 = FK5+HIP        & 0.35 & 0.33 & 0.35 & 0.34\\[4.0ex]
ratio of HIPPARCOS& 2.3\hphantom{0} &
                    1.9\hphantom{0} & 1.9\hphantom{0} &
                    1.8\hphantom{0}\\
to FK6 errors\\[0.5ex]\hline
\end{tabular}
\end{table}
%%%%%%%%%%%%%%%%%%%%%%%%%%%%%%%% End of Table 2 %%%%%%%%%%%%%%%%%%%%%%%%%%%%%

%%%%%%%%%%%%%%%%%%%
% Begin of        %
%                 %
% T a b l e  3    %
%                 %
%%%%%%%%%%%%%%%%%%%
\begin{table*}[th]
\caption
{
Error budget for GC+HIP proper motions in the `single-star mode'
}
\begin{tabular}{lcccccccc}
\hline\\[-1.0ex]
\multicolumn{9}{c}{Typical mean errors of proper motions}\\
\multicolumn{9}{c}{(in one component, averaged over $\mu_{\alpha\ast}$ and
$\mu_\delta$; units: mas/year)}\\[1.5ex]\hline\\[-2.0ex]
Sample of stars: & \multicolumn{2}{c}{29\,717 GC} &
\multicolumn{2}{c}{11\,773 GC} &
\multicolumn{2}{c}{1534 FK from GC} &
\multicolumn{2}{c}{1201 FK from GC}\\
Median of m$_V$: & \multicolumn{2}{c}{7\,\,.$\!\!\!^m$0} &
\multicolumn{2}{c}{6\,\,.$\!\!\!^m$8} &
\multicolumn{2}{c}{4\,\,.$\!\!\!^m$8} &
\multicolumn{2}{c}{4\,\,.$\!\!\!^m$9}\\
\hline\\[-0.5ex]
& rms average & median & rms average & median & rms average &
median & rms average & median\\[0.5ex]\hline\\[-1.3ex]
HIPPARCOS  & 1.47  & 0.73 & 0.75 & 0.69 & 0.82 & 0.63 & 0.68 & 0.61\\[1.5ex]
GC\\
\hspace*{0.5cm}random & 10.57\hphantom{1} & 9.38 & 8.59 & 7.55 & 3.33 & 2.31 &
3.44 & 2.45\\
\hspace*{0.5cm}system & 0.38 & 0.34 & 0.38 & 0.33 & 0.47 & 0.40 & 0.47 & 0.40\\
\hspace*{0.5cm}total & 10.58\hphantom{1} & 9.39 & 8.60 & 7.56 & 3.37 & 2.38 &
3.47 & 2.48\\[1.5ex]
$\mu_0$\\
\hspace*{0.5cm}random & 1.78 & 1.75 & 1.43 & 1.45 & 0.66 & 0.55 & 0.67 & 0.57\\
\hspace*{0.5cm}system & 0.15 & 0.11 & 0.12 & 0.11 & 0.19 & 0.14 & 0.18 & 0.14\\
\hspace*{0.5cm}total  & 1.78 & 1.75 & 1.43 & 1.45 & 0.66 & 0.55 & 0.67 &
0.57\\[4ex]
GC+HIP & 0.72 & 0.63 & 0.62 & 0.58 & 0.42 & 0.39 & 0.41 & 0.39\\[4ex]
ratio of HIPPARCOS & 2.0\hphantom{-} & 1.16 & 1.21 & 1.19 & 2.0\hphantom{-} &
1.6\hphantom{-} & 1.7\hphantom{-} & 1.6\hphantom{-}\\
to GC+HIP errors\\[0.5ex]\hline
\end{tabular}
\end{table*}

%%%%%%%%%%%%%%%%%%%%%%%%%%%%%% End of Table 3 %%%%%%%%%%%%%%%%%%%%%%%%%%%%%

In Table 3, the error budget for the combination GC+HIP of the GC data with
the HIPPARCOS observations are shown. The samples of stars are the following:
(a) \, all the 29\,717 GC stars which we have identified in the HIPPARCOS
Catalogue; (b) \, 11\,773 stars in the GC which have standard solutions by
HIPPARCOS (i.e., they do not occur in the DMS Annex of the HIPPARCOS Catalogue,
which would indicate a double or multiple object) and which are not known to be
double from ground-based observations and which have accurate values of
$\mu_{0 (GC)}$ (mean errors smaller than 2.0 mas/year in each coordinate); (c)
\, all the 1534 basic FK5 stars in the GC (one basic FK5 star is not in the
GC); (d) \, 1201 `apparently single' basic FK5 stars in the GC. Table 3 shows
that for the 11\,773 apparently single GC stars the combination GC+HIP produces
proper motions which are more accurate than the HIPPARCOS proper motions by
about 20\,\%. The overwhelming part of this improvement stems from the proper
motion $\mu_{0 (GC)}$. For the brighter GC stars, represented by the FK stars
in Table 3, the gain in accuracy is much larger, typically by a factor 1.5 to
2\,. A comparison of the results for GC+HIP given in Table 3 with those for
FK6\,=\,FK5+HIP in Table 2 for the basic FK5 stars confirms that the older
ground-based observations carry a very large weight in the combination with
the HIPPARCOS data: For the basic FK5 stars, the mean errors of the proper
motions of the combination GC+HIP are only slightly larger (by about 20\,\%)
than those of the FK6\,=\,FK5+HIP.

The error budgets for the stars in the bright and faint extension of the FK5
and for the remaining FK4Sup stars are not given here. For these stars the gain
in accuracy of their proper motions by using our combination method lies
inbetween the gain for the basic FK5 stars and that for the GC stars.

We would like to warn those readers who would like to test our combination
method with the help of the numbers given in
Tables 2 and 3, using the equations
given in Sect. 2\,. The quantities given in
Tables 2 and 3 are ensemble averages
of the mean errors $\varepsilon_{\mu}$, while our combination method works
individually star by star. Since $< 1/\varepsilon^2_{\mu} >$ differs in general
from $1/< \varepsilon^2_{\mu} >$, such a test is not meaningful. A numerical
check of our method can, of course, be carried out by using the numbers given
in Table 1 for the individual star $\alpha$ Ari.

It should be emphazised that the gain in accuracy is significantly larger in
the `long-term prediction mode' (Wielen et al. 1998) in which we take into
account the `cosmic errors' in the HIPPARCOS results (see Sect. 4). The error
budget for the long-term predictions will be discussed in detail in a
subsequent paper.

%Section 7
\section{Conclusions and outlook}

We have shown that the combination of ground-based compilation catalogues with
the HIPPARCOS data produces for many stars proper motions which are
significantly more accurate than the HIPPARCOS proper motions themselves. For
stars which are bright ($\sim 5^m$), well-observed and truly single, the gain
in accuracy is about a factor of two. For stars of intermediate magnitudes
($\sim 7^m$), the smaller gain of about 20\,\% on average is still valuable for
many applications.

The combination method described above has been used intensively and
successfully in the construction of the FK6, in which the basic FK5 is combined
with HIPPARCOS. The method described here provides the FK6 results in the
`single-star mode' (see Wielen et al. 1998). The first part of the FK6 will be
published in the
near future. The second part of the FK6 deals with double stars
and requires a slightly modified version of our combination method.

In order to take into account the cosmic errors in the HIPPARCOS results, which
are caused by undetected astrometric binaries, our combination method has to be
changed slightly. The corresponding procedures, which lead to the `long-term'
and `short-term' prediction modes (Wielen et al. 1998, 1999),
will be presented
in a subsequent paper.

%\begin{acknowledgements}
%T e x t
%\end{acknowledgements}

%\listofobjects


\begin{thebibliography}{}

\bibitem[]{}
Bien  R., Fricke  W., Lederle  T., Schwan  H., 1978, Ver\"off. Astron.
Rechen-Inst. Heidelberg No. 29

\bibitem[]{}
Boss  B., Albrecht  S., Jenkins  H., Raymond  H., Roy  A.J., Varnum
W.B., Wilson  R.E., 1937, General Catalogue of 33\,342 Stars for the Epoch
1950, Carnegie Institution of Washington, Publ. No. 486

\bibitem[]{}
Eichhorn  H., 1974, Astronomy of Star Positions, Frederick Ungar Publ.
Co., New York

\bibitem[]{}
ESA  1997, The Hipparcos Catalogue, ESA SP-1200

\bibitem[]{}
Fricke  W., Schwan  H., Lederle  T., Bastian  U., Bien  R., Burkhardt  G., du
Mont  B., Hering  R., J\"ahrling  R., Jahrei{\ss}  H., R\"oser  S.,
Schwerdt\-feger  H.M.,
Walter  H.G., 1988, Ver\"off. Astron. Rechen-Inst.
Heidelberg No. 32

\bibitem[]{}
Fricke  W., Schwan  H., Corbin  T., Bastian  U., Bien  R., Cole  C., Jackson
E., J\"ahrling  R., Jahrei{\ss}  H., Lederle  T., R\"oser  S., 1991, Ver\"off.
Astron. Rechen-Inst. Heidelberg No. 33

\bibitem[]{}
Kopff  A., 1937, Ver\"off. Astron. Rechen-Inst. Berlin No. 54

\bibitem[]{}
Kopff  A., 1938, Abh. Preu{\ss}. Akad. Wiss., Jahrgang 1938, Phys.-math.
Klasse, No. 3

\bibitem[]{}
Kopff  A., Nowacki  H., Strobel  W., 1964, Ver\"off. Astron. Rechen-Inst.
Heidelberg No. 14

\bibitem[]{}
Schwan  H., Bastian  U., Bien  R., J\"ahrling  R., Jahrei{\ss}  H., R\"oser
S., 1993, Ver\"off. Astron. Rechen-Inst. Heidelberg No. 34

\bibitem[]{}
Wielen  R., 1988,
In: IAU Symposium No. 133,  Mapping the Sky, eds.
Debarbat S., Eddy J.A., Eichhorn H.K., Upgren A.R.,
Kluwer Publ. Comp., Dordrecht,
p. 239

\bibitem[]{}
Wielen  R., 1995a, A\&A 302, 613

\bibitem[]{}
Wielen  R., 1995b,
In: Future Possibilities for Astrometry in Space, eds.
Perryman M.A.C., van Leeuwen F., ESA SP-379, p. 65

\bibitem[]{}
Wielen  R., 1997, A\&A 325, 367

\bibitem[]{}
Wielen  R., 1998, T\"atigkeitsbericht des Astron. Rechen-Inst.
Heidelberg f\"ur das Jahr 1997 = Mitt. Astron. Ges. No. 81, 327

\bibitem[]{}
Wielen  R., Schwan  H., Dettbarn  C., Jahrei{\ss}  H., Lenhardt  H., 1997,
In:
Hipparcos Venice '97, Presentation of the Hipparcos and Tycho Catalogues
and first astrophysical results of the Hipparcos space astrometry mission, eds.
Battrick  B., Perryman  M.A.C., Bernacca  P.L., ESA SP-402, p. 727

\bibitem[]{}
Wielen  R., Schwan  H., Dettbarn  C., Jahrei{\ss}  H., Lenhardt  H., 1998,
In:
The Message of the Angles -- Astrometry from 1798 to 1998, Proceedings of the
International Spring Meeting of the
Astronomi\-sche Gesellschaft, held in Gotha,
11-15 May 1998, eds. Brosche P., Dick W.R., Schwarz O., Wielen R.,
Acta
Historica Astronomiae, Vol. 3, Verlag Harri Deutsch, Thun and Frankfurt am
Main, p. 123

\bibitem[]{}
Wielen  R., Schwan  H., Dettbarn  C., Jahrei{\ss}  H., Lenhardt  H., 1999,
In:
Modern Astrometry and Astrodynamics,
Proceedings of the International Conference
honouring Heinrich Eichhorn,
held at Vienna Observatory, Austria, 25-26 May
1998,
eds. Dvorak R., Haupt H.F., Wodnar K.,
Verlag der \"Osterreichischen Akademie der
Wissenschaften, Wien, p. 161
\end{thebibliography}
\end{document}